\documentclass[aps,prl,twocolumn,groupedaddress]{revtex4-1}

\newcommand{\la}{\langle}
\newcommand{\ra}{\rangle}
     		  \usepackage{graphicx}
     		  \usepackage{epstopdf}
\usepackage{color}
\usepackage{float}
\usepackage{amsfonts}
\usepackage{amsmath}
\usepackage{mathrsfs}
\usepackage[toc]{appendix}
\usepackage{graphicx}
\usepackage{amsthm}
\usepackage{amssymb}
\makeatletter
\def\NAT@def@citea{\def\@citea{\NAT@separator}}
\makeatother

\begin{document}


\title{Message passing and moment closure for susceptible-infected-recovered epidemics on finite networks}


\author{Robert R. Wilkinson, Kieran J. Sharkey}
\affiliation{The University of Liverpool}


\date{\today}

\begin{abstract}
The message passing approach of Karrer and Newman [Phys. Rev. E \textbf{82}, 016101 (2010)] is an exact and practicable representation of susceptible-infected-recovered dynamics on finite trees. Here we show that, assuming Poisson contact processes, a pair-based moment closure representation [Sharkey, J. Math. Biol. \textbf{57}, 311 (2008)] can be derived from their equations. We extend the applicability of both representations and discuss their relative merits. On arbitrary time-independent networks, as was shown for the message passing formalism, the pair-based moment closure equations also provide a rigorous lower bound on the expected number of susceptibles at all times.   
\end{abstract}

\pacs{89.75.Hc,87.23.Ge,87.10.Ed}

\maketitle



\section{Introduction}

There is a large, rapidly growing body of research relating to `network epidemiology', where real-world host-to-host disease dynamics are modelled as stochastic processes that take place upon contact networks \cite{keeldan}. In many other areas, including transportation, logistics and social interactions, stochastic processes are also propagated across networks \cite{newnet}. These random processes which act directly on network nodes, or across network links, alter the states of the individuals in the network. Finding mathematical descriptions for the time evolution of the probabilities of the states of the individuals is essential to understanding the dynamics of these systems and how to control them. Recently, two practicable and provably exact representations of stochastic epidemic models on finite trees have been proposed. These are the message passing approach of Karrer and Newman \cite{karnew} and the pair-based moment closure representation of Sharkey \emph{et. al.} \cite{sharkey2008,sharkey2011,sharkeyA}. Here we generalise and compare these two representations, and show that the pair-based equations can be derived from the message passing formalism. 


\section{susceptible-infected-recovered dynamics on tree networks}

We define an arbitrary network to be a directed graph (digraph) $D=(V,A)$, where $V$ represents the set of all individuals in the population and the existence of an arc $(i,j) \in A$ ($i,j \in V$) represents the ability of $i$ to make contacts to $j$. The population size is denoted by $N=|V|$. Following Karrer and Newman \cite{karnew}, while allowing heterogeneity in the ordered-pairwise interactions, we define the functions $f_{ij}(\tau)$ $(\forall i,j \in V : (j,i) \in A)$ such that the probability that $j$ makes an infectious contact to $i$ within time period $t$ of being infected is given by $\int_0^t f_{ij}(\tau) \mbox{d} \tau$. If $i \in V$ is susceptible when it receives an infectious contact then it immediately becomes infected. The time that it takes $i$ to become recovered (infectious contacts from $i$ permanently cease), after it has been infected, is distributed according to an individual-specific probability density function $r_i(\tau)$. We will also assume that $f_{ij}(\tau)=h_{ij}(\tau) \int_{\tau}^{\infty} r_j(\tau') \mbox{d}\tau'$, where $\int_0^t h_{ij}(\tau) \mbox{d}\tau$ is the probability that $j$ makes any contact to $i$ within time period $t$ of being infected. The above defines susceptible-infected-recovered (SIR) dynamics on an arbitrary network. Throughout, we assume that the initial states of individuals are statistically independent.

In this section we consider SIR dynamics on the full class of digraphs where the underlying graphs are trees or forests. We refer to such digraphs as tree networks.

\subsection{Message passing formalism for individuals}

The fundamental quantity in the message passing formalism for tree networks is $H^{i \leftarrow j}(t)$ (also defined for non-tree networks), where $(j,i) \in A(D)$, which is the probability that $i$ has not received an infectious contact from $j$ by time $t$ given that $i$ is in the cavity state (contacts \emph{from} $i$ are not permitted). Note that placing an individual in the cavity state does not affect its fate since its ability to contact others only comes into play if it is infected. Here we will allow heterogeneity in the contact and recovery processes and initial conditions, and also allow individuals to be vaccinated. Thus we can express the message passing equation as
\begin{eqnarray}  \nonumber
H^{i\leftarrow j}(t)&=&1- \int_0^t f_{ij}(\tau)\Big[1-y_j- z_j \Phi^j_i(t-\tau)\Big] \mbox{d} \tau , \\   
&&
\label{H1}
\end{eqnarray}
where $y_j$ and $z_j$ are the probabilities that $j$ is recovered at $t=0$ (vaccinated) and susceptible at $t=0$ respectively ($j$ is initially infected with probability $1-y_j-z_j$) and $\Phi_i^j(t)$ is the probability that $j$ has not received an infectious contact by time $t$ given that $i$ and $j$ are both in the cavity state. For the tree networks considered in the present section, we can substitute in Eq. \ref{H1} \cite{karnew}: 
\begin{eqnarray}
\label{phi1}
\Phi_i^j(t)=\prod_{ \substack{k \in N_j \\  k \neq i}} H^{j \leftarrow k}(t) ,
\end{eqnarray}
 where $(j,i) \in A(D)$ and $N_j$ is the set of individuals (neighbours) from which there are arcs towards $j$. If $\nexists k: k \in N_j,k \neq i$, then we define the right-hand side of Eq. 2 to be equal to 1. 
Here, $\Phi_i^j(t)$ can be expressed as a product of probabilities of independent events because of the simple structure of tree networks, i.e. there is no more than one simple directed path from any individual to any other, and no cycles. This is discussed in more detail in the subsequent section on non-tree networks. 

Equation \ref{H1} can now in principle be solved (at every point in time) for every $i,j \in V : j \in N_i$ via the associated system of integral equations, as discussed in \cite{karnew} (the number of these equations is equal to $|A(D)|$). It is then straightforward to obtain the time evolution of the probability of an individual being in a particular state \cite{karnew}:
\begin{eqnarray} \nonumber
\label{singles}
\langle S_i \rangle &=& z_i\prod_{j \in N_i} H^{i \leftarrow j}(t), \\ \nonumber
\la R_i \ra&=& y_i + \int_0^t r_{i} (\tau) \Big[1-y_i-\la S_i \ra_{t-\tau} \Big] \mbox{d}\tau, \\  
\la I_i  \ra &=& 1 - \la S_i \ra - \la R_i \ra,
\end{eqnarray}
where $\la A_i \ra$ and $\la A_i  \ra_{t-\tau}$ are the probabilities that $i$ is in state $A \in \{S,I,R\}$ (susceptible, infected or recovered) at time $t$ and at time $t-\tau$ respectively. Note that the expected number of individuals in state $A$ at time $t$ is given by $\sum_{i \in V} \la A_i \ra$.

By setting $z_i,y_i \in \{0,1\}$ $\forall i\in V$, we can consider any pure initial system 
state (note that we assume initially infected individuals `become' infected at $t=0$). In this case, we could eliminate $y_i$ from our equations by removing from the network those individuals that are vaccinated. However, we can also consider mixed (probabilistic) initial system states, as long as the states of individuals are initially independent. For instance, we might consider the case where every individual is independently vaccinated with probability $y$. 

 We obtain the specific form in the Karrer and Newman article by setting $f_{ij}=f,r_i=r,z_i=z,y_i=0$ $\forall i,j \in V :j \in N_i $. In this case, the solution represents a measure of an `average epidemic' but we note that the initial distribution for the total number of infecteds is binomial and includes the possibility of no initial infecteds. Typically, we are more interested in the expected outcome when a single initial infected individual is seeded uniformly at random in the population (see for example \cite{meyers}). We identify two methods for computing this (appendix A).

 \subsection{Message passing formalism for pairs}

To start to link the message passing method with the pair-based models \cite{sharkey2008,sharkey2011,sharkeyA}, we express some relevant probabilities for connected pairs in tree networks. The probability $\langle S_iS_j \rangle$ that neighbouring individuals $i$ and $j$ ($(i,j) \in A$ or $(j,i)\in A$) are susceptible at time $t$ is given by
 \begin{eqnarray}
 \label{pair1}
 \langle S_iS_j \rangle = z_i z_j \Phi^i_j(t) \Phi^j_i(t) ,
 \end{eqnarray}
which follows from the fact that this pair state can only be destroyed by the infection of $i$, not via $j$, or the infection of $j$, not via $i$.

The probability $\la S_i I_j \ra$ that $i$ is susceptible and $j$ is infectious at time $t$ (in a tree network) is more difficult to formulate (here there is an arc from $j$ to $i$). Firstly, we place both $i$ and $j$ in the cavity state, with the exception that we allow infectious contacts from $j$ to $i$. Now, we know that for this pair state to occur we need $i$ to be susceptible at $t=0$ and for it not to receive an infectious contact from any of its neighbours (ignoring $j$ for the moment) by time $t$; the probability of this is $z_i \Phi^i_j(t)$. Secondly, we multiply this quantity by the probability $P_{ij}(t)$ of the independent event that $j$ is initially susceptible and receives an infectious contact from a neighbour other than $i$ (or is infected at $t=0$) and then remains infectious, without making an infectious contact to $i$, until time $t$. Thus: 
\begin{eqnarray}  
\label{pair21}
\la S_i I_j \ra &=&    z_i \Phi^i_j(t)P_{ij}(t) .
\end{eqnarray}
However, $1-P_{ij}(t)$ can be computed as a sum of probabilities of mutually exclusive events, i.e. $j$ is initially recovered or vaccinated (with probability $y_j$), or $j$ is initially susceptible and does not receive an infectious contact (with probability $z_j \Phi^j_i(t)$), or $j$ does receive an infectious contact (or is initially infected) but then also makes an infectious contact to $i$ or recovers without making such a contact, by time $t$. Therefore, we now have:
\begin{eqnarray}  \nonumber
\label{pair2}
\la S_i I_j \ra &=&    z_i \Phi^i_j(t)\bigg{[} 1 - y_j -  z_j\Phi^j_i(t) \\ \nonumber
&& -  \int_0^t \Big( f_{ij}(\tau)+g_{ij}(\tau) \Big)  \\  
&& \times \Big(1-y_j-z_j \Phi^j_i(t-\tau)\Big)  \mbox{d}\tau \bigg{]} ,
\end{eqnarray}
where $g_{ij}(\tau)=r_j(\tau) \int_{\tau}^{\infty} h_{ij}(\tau') \mbox{d} \tau'$ is defined such that $\int_0^t g_{ij}(\tau) \mbox{d}\tau$ is the probability that $j$ recovers from the infection within time period $t$ of being infected, without contacting $i$ during this period.

\subsection{Equations for Poisson contact processes}

For the remainder of this section, we consider a special case of the message passing formalism where the contact processes are Poisson, i.e. $f_{ij}(\tau)= T_{ij}e^{-T_{ij}\tau } \int_\tau^\infty r_j(\tau') \mbox{d} \tau'$ where $T_{ij}$ is the rate at which $j$, when it is infected, makes infectious contacts to $i$. 
 In this case, it can be shown (see appendix B) that a deterministic time series for $\la A_i \ra, \forall A \in \{S,I,R\}, \forall i \in V$, can in principle be obtained by integrating the following integro-differential equations, with $H^{i \leftarrow j}(0)=1$ $(\forall i,j \in V : j \in N_i)$:
\begin{eqnarray}  \nonumber
\label{genmess}
\dot{H^{i \leftarrow j}(t)}&=&-T_{ij} \bigg{[}  H^{i \leftarrow j}(t) - y_j - z_j\Phi^j_i(t)  \\ \nonumber 
&& - \int_0^t r_j(\tau) e^{-T_{ij}\tau} \Big(1-y_j-z_j\Phi^j_i(t-\tau)\Big)  \mbox{d}\tau \bigg{]} , \\
&&
\end{eqnarray}
and then making use of Eq. \ref{singles}.

 The above message passing system (Eqs. \ref{singles} and \ref{genmess}), in conjunction with Eqs. \ref{pair1} and \ref{pair2} for the states of connected pairs, implies the following pair-based system of integro-differential equations (see appendix B):
\begin{widetext}
\begin{eqnarray} \nonumber
\label{genpair}
\dot{\la S_i\ra}&=&-\sum_{j \in N_i} T_{ij}\la S_iI_j\ra, \qquad \dot{\la I_i\ra}=\sum_{j \in N_i} T_{ij}\la S_iI_j\ra -  \int_0^t  \sum_{j \in N_i} T_{ij}\la S_iI_j\ra_{t-\tau} r_i(\tau) \mbox{d} \tau - (1-y_i-z_i)r_i(t),  \\ \nonumber 
\dot{\la S_iI_j\ra}&=&\sum_{\substack{k \in N_j \\  k\neq i}} T_{jk}\frac{\la S_iS_j\ra\la S_jI_k\ra}{\la S_j\ra}-\sum_{\substack{k \in N_i \\ k\neq j}} T_{ik}\frac{\la S_iI_k\ra\la S_iI_j\ra}{\la S_i\ra} -T_{ij}\la S_iI_j\ra  \\ \nonumber
&&-  \int_0^t \sum_{\substack{k \in N_j \\ k \neq i}} T_{jk}  \frac{\la S_iS_j\ra_{t-\tau} \la S_jI_k\ra_{t-\tau}}{\la S_j\ra_{t-\tau}} r_j(\tau)e^{-T_{ij}\tau} \frac{\Phi^i_j(t)}{\Phi^i_j(t-\tau)} \mbox{d}\tau  - z_i(1-y_j-z_j)r_j(t)e^{-T_{ij}t}\Phi^i_j(t),  \\  
\dot{\la S_iS_j\ra}&=&-\sum_{\substack{k \in N_i \\ k\neq j}}T_{ik}\frac{\la S_i I_k\ra\la S_iS_j\ra}{\la
S_i\ra}-\sum_{\substack{k \in N_j \\ k\neq i}}T_{jk}\frac{\la S_iS_j\ra\la S_jI_k\ra}{\la S_j\ra}, \qquad \dot{\Phi^i_j(t)}=- \sum_{\substack{k \in N_i \\ k \neq j}} T_{ik}\frac{\la S_i I_k  \ra}{\la S_i \ra} \Phi^i_j(t) ,  \\   \nonumber
\end{eqnarray}
\end{widetext}  
and $\la R_i \ra=1-\la S_i \ra - \la I_i \ra$. Note that $\Phi^i_j(t)$ can be expressed as
\begin{eqnarray} 
\label{phi}
\Phi^i_j(t)&=&\mbox{exp}\left(-\int_0^t  \sum_{\substack{k \in N_i \\ k \neq j}} T_{ik}\frac{\la S_i I_k  \ra_{t'}}{\la S_i \ra_{t'}} \mbox{d}t' \right)  ,
\end{eqnarray}
and $\la S_i S_j \ra$ can be expressed in terms of $\Phi^i_j(t)$ (Eq. \ref{pair1}).
We recognise Eq. \ref{genpair} as a generalisation (to arbitrary recovery processes) of the pair-based moment closure equations~\cite{sharkeyA} which assume the following approximation for the probability of the state of a connected triple: $\la A_iB_jC_k\ra=\la A_iB_j\ra\la B_jC_k\ra/\la B_j\ra$ where $\la B_j\ra \neq 0$. In the case of SIR dynamics on tree networks we know that this approximation is exact when $B=S$ (susceptible), as in Eq. \ref{genpair} \cite{sharkeyA,kissA}.   
It is clear that the number of equations will be of the order of the number of directed links in the network, i.e. $|A(D)|$. 

These systems are exact for any tree network. More specifically, they are exact for any digraph where there is no more than one simple directed path from any individual to any other individual, and no cycles. The reason for this will be made clear in the later section on non-tree networks. Indeed, for some non-tree networks the equations are still exact or may become exact for certain initial conditions, as discussed by Sharkey \emph{et al.} \cite{sharkeyA} in relation to the pair-based representation for Markovian SIR dynamics.

The message passing system (Eqs. \ref{singles} and \ref{genmess}) and the pair-based system (\ref{genpair}) are equivalent in the sense that they produce the same time series for the probabilities of the states of individuals. Here, we shall consider both systems for exponential and fixed recovery processes.

\subsubsection{Exponential infectious periods}

For exponential recovery we have $r_i(\tau)=\gamma_i e^{-\gamma_i \tau}$ where $\gamma_i$ is an individual-specific constant. Substituting this into Eq. \ref{genmess}, we obtain a system of ordinary differential equations (ODEs) which are straightforward to integrate numerically:
\begin{eqnarray}  \nonumber
\label{exp}
\dot{H^{i \leftarrow j}(t)}&=&  \gamma_j\Big[1-H^{i \leftarrow j}(t)\Big]  \\ 
&&-T_{ij}\Big[ H^{i \leftarrow j}(t)-y_j -  z_j \Phi^j_i(t) \Big].  \\  \nonumber
\end{eqnarray}
The time evolution of the probability of an individual being in a particular state is then obtained via Eq. \ref{singles} with the exception that we can now conveniently obtain the probability of an individual being recovered via the differential equation:
\begin{eqnarray} 
\label{expsing}
\dot{\la R_i \ra}&=& \gamma_i \la I_i \ra .
\end{eqnarray}

After computing the time derivatives of the probabilities of an individual being in each of the possible states (using Eqs. \ref{singles} and \ref{exp}), and expressing the results in terms of individuals and pairs (Eqs. \ref{singles} to \ref{pair2}), the pair-based moment closure system of Sharkey \emph{et al.} \cite{sharkeyA} can be derived. Alternatively, it can be derived from the more general pair-based system (\ref{genpair}) via the substitution $r_i(\tau)=\gamma_i e^{-\gamma_i \tau}$ and making use of Eqs. \ref{singles} to \ref{pair2}. 

\subsubsection{Fixed infectious periods}

\begin{figure}
 \centerline{\includegraphics[width=0.5\textwidth]{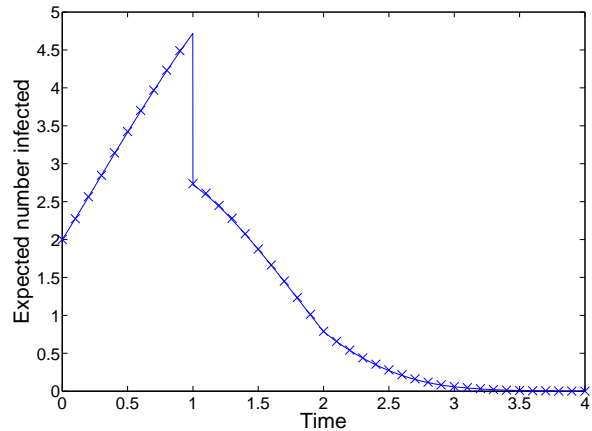}}
\caption{Here we investigate SIR dynamics (Poisson contact, fixed recovery) on an undirected tree network of 10 individuals. We set the infectious period to unity for all individuals and $T_{ij}=1$ $ \forall i,j \in V : j \in N_i$. Two non-adjacent index-individuals were selected to be initial infecteds, while each non-index individual was vaccinated with probability 1/10 and susceptible otherwise (at $t=0$). The line represents the output from our representation (Eqs. \ref{fix} and \ref{fix2}) while the crosses indicate corresponding numerical results from 10,000 full stochastic simulations.}
\label{F1}
\end{figure}

For fixed infectious periods (or fixed time to recovery) we have $r_i(\tau)=\delta(\tau-\omega_i)$ where $\delta(\tau)$ is the Dirac $\delta$ function and $\omega_i$ is the time it takes $i$ to recover once it has been infected. Substituting this into Eq. \ref{genmess}, we obtain a system of delay differential equations (DDEs) which are straightforward to integrate numerically:
\begin{eqnarray}  \nonumber
\label{fix}
\dot{H^{i \leftarrow j}(t)}&=&-T_{ij} \bigg{[}  H^{i \leftarrow j}(t) - y_j - z_j\Phi^j_i(t)  \\ \nonumber
&&- \theta (t-\omega_j) e^{-T_{ij}\omega_j}\Big(1 - y_j - z_j \Phi^j_i(t - \omega_j) \Big) \bigg{]}, 
     \\ 
&&
\end{eqnarray}
where $\theta(t)$ is the Heaviside step function. For the individuals we have:
\begin{eqnarray} \nonumber
\label{fix2}
\la S_i \ra &=& z_i \prod_{j \in N_i}H^{i \leftarrow j}(t),  \\  \nonumber
\la R_i \ra&=& y_i + \theta (t-\omega_i)\Big[1-y_i- \la S_i \ra_{t - \omega_i}\Big],  \\  
\la I_i \ra & = & 1- \la S_i \ra - \la R_i \ra.
\end{eqnarray}
Similarly, a DDE version of system~\ref{genpair} can also be derived. Fig \ref{F1} matches the output from Eqs. \ref{fix} and \ref{fix2} to stochastic simulation for a tree network of 10 individuals, illustrating exactness. 

\section{SIR dynamics on non-tree networks}

Karrer and Newman \cite{karnew} proved that their message passing formalism, when applied to non-tree networks, provides a rigorous lower bound on $\la S_i \ra$ $\forall i \in V$. Here we repeat their analysis in order to confirm that this bound is still obtained in our slightly more general setting, and to show that the pair-based system (\ref{genpair}) consequently provides the same bound on $\la S_i \ra$ $ \forall i \in V$.

Following Karrer and Newman, we represent SIR dynamics on an arbitrary digraph $D=(V,A)$ by randomly weighting and removing the arcs as follows: 1) assign an infectious period $\tau_i$ to every individual $i\in V$, sampling from $r_i$. 2) weight every arc $(j,i) \in A$ with a contact time $\omega_{ij}$, sampling from contact functions $h_{ij}$. 3) for every arc $(j,i) \in A$, if its weighting $\omega_{ij}>\tau_j$ then completely remove it from the digraph. 4) for every individual $i \in V$, with probability $y_i$, completely remove every arc emanating from $i$.

 The resulting weighted digraph is denoted $D'$. $n_{iB}(D')$, where $B \subseteq N_i$, denotes the set of individuals from which $i$ can be reached by a simple directed path of total weighting less than $t$, such that a member of $B$ is the penultimate individual, given that $i$ is in the cavity state (all arcs emanating from $i$ are removed).

Let $i \leftarrow B$, where $B \subseteq N_i$, denote the event that $i$ (in the cavity state) does not receive any infectious contacts from any of the members of $B$ by time $t$. Let $|N_i|=M$ and let us label each of these neighbours as $N_i^{(1)},N_i^{(2)}, \ldots, N_i^{(m)}, \ldots , N_i^{(M)}$ where the ordering is arbitrary. We can now express $\la S_i \ra$ as a product of conditional probabilities:
\begin{eqnarray} \nonumber
\label{condprod}
\la S_i \ra & =& z_i P ( i \leftarrow \cup_{p=1}^M N_i^{(p)} ) \\ \nonumber
&=& z_i P ( i \leftarrow N_i^{(1)} ) \\ \nonumber
&& \times P ( i \leftarrow N_i^{(2)} \mid i \leftarrow N_i^{(1)} ) \\ \nonumber
&& \times \ldots \times  P ( i \leftarrow N_i^{(m)}  \mid i \leftarrow \cup_{p=1}^{m-1}N_i^{(p)}  ) \\ \nonumber
&& \times \ldots \times P ( i \leftarrow N_i^{(M)} \mid i \leftarrow \cup_{p=1}^{M-1}N_i^{(p)} ). \\
\end{eqnarray}

 The particular way in which $D'$ is constructed means that, for any $j \in N_i$, we have:
\begin{equation}
\label{expec}
P(i \leftarrow j)= \mathbb{E} \left[ \prod_{k \in n_{i j}} \frac{z_k}{1-y_k}  \right],
\end{equation}
where the expectation operator is here applied to a function of the random weighted digraph of which $D'$ is a single realisation, and the product is assumed to be equal to 1 when $n_{ij}=\emptyset$. Equation \ref{expec} follows from the fact that all members of $n_{ij}(D')$ must be initially susceptible if $D'$ is to represent the event that $i$ (in the cavity state) does not receive an infectious contact from $j$ by time $t$. $z_k /( 1 - y_k)$ is the probability that $k$ is initially susceptible given that it is not vaccinated (we excluded the possibility of a member of $n_{ij}(D')$ being vaccinated in step 4 of the construction of $D'$). Similarly, for any $B \subset N_i : j \notin B$, we can write:
\begin{equation}
P(i \leftarrow j \mid i \leftarrow B)= \mathbb{E} \left[ \prod_{k \in n_{i j} \setminus n_{iB}} \frac{z_k}{1-y_k}  \right]. 
\end{equation}

Now, since $n_{i j} \setminus n_{iB} \subseteq n_{i j}$, with set equality occurring for tree networks, we have:
\begin{equation}
\label{ind}
P(i \leftarrow j) \le P(i \leftarrow j \mid i \leftarrow B),
\end{equation}
with equality occurring for tree networks. In fact, $n_{ij}(D')$ and $n_{iB}(D')$ are necessarily disjoint sets if there is no more than one simple directed path from any individual to any other individual in $D$. 

Inequality \ref{ind} implies that the conditioning in each term of the product in Eq. \ref{condprod} can only serve to increase the total probability. Therefore
\begin{equation} 
\la S_i \ra  \ge  z_i \prod_{j \in N_i} P(i \leftarrow j) = z_i \prod_{j \in N_i} H^{i \leftarrow j} (t). 
\end{equation}

Inequality \ref{ind} also implies that
\begin{equation}
\label{phi2}
\Phi^j_i(t) \ge \prod_{\substack{k \in N_j \\  k \neq i}} P(j \leftarrow k \mid i \mbox{ in cavity}),
\end{equation}
where we have ignored $P(j \leftarrow i \mid i \mbox{ in cavity})$ since it is necessarily equal to 1 ($\Phi_i^j(t)$ is the probability that $j$ has not received an infectious contact by time $t$ given that $i$ and $j$ are both in the cavity state, where $(j,i) \in A(D)$ - see Eq. \ref{H1}). Now, taking $i$ out of the cavity state, we only increase (or leave the same) the probability of infectious contact across any arc (replacing the arcs emanating from $i$ cannot result in $n_{jk}$ losing any of its members), and so
\begin{equation}
\label{cav}
 \prod_{\substack{k \in N_j \\  k \neq i}}  P(j \leftarrow k \mid i \mbox{ in cavity}) \ge \prod_{\substack{k \in N_j \\  k \neq i}}H^{j \leftarrow k}(t), 
\end{equation}
with equality occurring for tree networks. Notice that this, in conjunction with equality in \ref{ind} and \ref{phi2}, implies Eq. \ref{phi1} for tree networks. However, we also get equality in \ref{cav} whenever there are no cycles in the digraph $D$. Therefore, sufficient requirements for the exactness of Eqs. \ref{phi1} to \ref{genpair} are that (1) there is no more than one simple directed path from any individual to any other individual in $D$ and (2) there are no cycles in $D$. Equations \ref{H1}, \ref{phi2} and \ref{cav} imply that
\begin{equation}
\label{ineqH}
H^{i\leftarrow j}(t) \ge 1- \int_0^t f_{ij}(\tau)\Big[1-y_j- z_j \prod_{\substack{k \in N_j \\  k \neq i}}H^{j \leftarrow k} (t-\tau)\Big] \mbox{d} \tau.  
\end{equation}

Following Karrer and Newman \cite{karnew}, we define the function:
\begin{equation}
\label{F}
F^{i\leftarrow j}(t) = 1- \int_0^t f_{ij}(\tau)\Big[1-y_j- z_j \prod_{\substack{k \in N_j \\  k \neq i}}F^{j \leftarrow k} (t-\tau)\Big] \mbox{d} \tau,  
\end{equation}
and note that it corresponds to the way in which $H^{i \leftarrow j}(t)$ can be expressed for tree networks (Eqs. \ref{H1} and \ref{phi1}). Now, using the iterative procedure which they suggest (see appendix C), it can be shown that $H^{i \leftarrow j} (t) \ge F^{i \leftarrow j}(t)$ $(\forall i,j,t: j \in N_i) $. This means that
\begin{eqnarray}
\la S_i \ra & \ge& z_i\prod_{j \in N_i}H^{i \leftarrow j} (t) \ge z_i\prod_{j \in N_i}F^{i \leftarrow j} (t). 
\end{eqnarray}

For Poisson contact processes, the (approximate) dynamics can be cast as systems of differential equations in both formalisms (all occurrences of $H, S, I$ and $R$, in Eqs. \ref{H1} to \ref{genpair}, are changed respectively to $F, X, Y$ and $Z$ - indicating inexactness). Since they are implied by the message passing formalism, the solution of the pair-based equations (\ref{genpair}) on non-tree networks, i.e. arbitrary digraphs, provides a rigorous lower bound on $\la S_i \ra$ and approximations for $\la I_i \ra, \la R_i \ra$ $\forall i \in V$.

Figure 2 illustrates the application of the message passing approach to SIR dynamics with Poisson contact processes and fixed recovery processes on a non-tree network (we use Eqs. \ref{phi1}, \ref{fix} and \ref{fix2}, changing all occurrences of $H, S, I$ and $R$, to $F, X, Y$ and $Z$, respectively).

\begin{figure}
 \centerline{\includegraphics[width=0.5\textwidth]{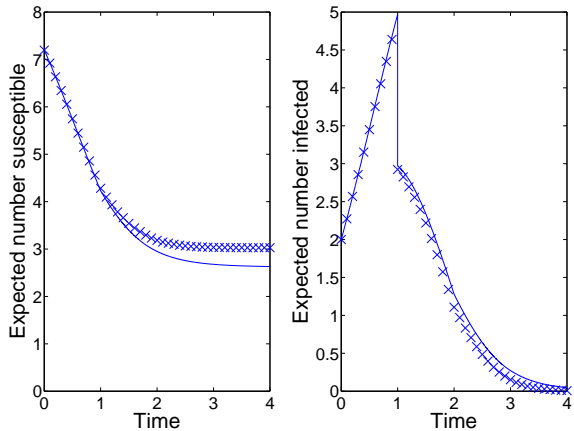}}
\caption{The same scenario as in Fig 1 except that two extra undirected connections, i.e. four arcs, have been added to the network, creating multiple cycles. The lines represent the output from our representation (Eqs. \ref{fix} and \ref{fix2}) while the crosses indicate corresponding numerical results from 10,000 full stochastic simulations.}
\label{F2}
\end{figure}

\section{Conclusions}

For Poisson contact processes, the message passing formalism can be cast as a system of integro-differential equations, which conveniently simplify to ODEs for exponential recovery processes and DDEs for fixed recovery processes. However, we note that for certain other biologically feasible sets of functions $\{ f_{ij}(\tau): i,j \in V , j \in N_i \}$, which do not correspond to Poisson contact processes, the message passing formalism may still allow the dynamics to be obtained via systems of ODEs or DDEs. See, for example, the `top hat' function discussed in \cite{karnew}. This is a clear advantage of the message passing formalism over the moment-closure formalism, the latter seeming to require the contact processes to be Poisson. In fact, for arbitrary contact and recovery processes, the message passing formalism is theoretically solvable as a system of integral equations.

Other advantages of the message passing approach are its applicability in the domain of random graph ensembles and, by considering $H^{i \leftarrow j}(t)$ (or $F^{i \leftarrow j}(t)$) $\forall i,j: j \in N_i$ in the limit as $t \to \infty$, its connection to percolation-based theory for final outcome statistics \cite{karnew}.


The pair-based formalism is a special case of the message passing approach in the sense that it seems to only apply to Poisson contact processes. In this case, the message passing system (\ref{genmess}) is more efficient than the pair-based system (\ref{genpair}) in terms of the number of equations. However, it is not immediately obvious how to extend the applicability of the message passing equations. For example, to generate exact equations for non-tree networks, or to susceptible-infected-susceptible dynamics. Conversely, the pair-based (moment closure) representation can allow both of these extensions in a straightforward way \cite{kissA,nagy}. Since the physical meaning of each term in the pair-based system is clear, it is also straightforward to make this system applicable to multiple competing diseases on the same network - the number of equations then grows linearly with the number of diseases. However, we note that in the context of configuration network ensembles and Poisson contact and recovery processes, Miller \cite{miller} has shown that the dynamics for competing diseases can be solved via a low-dimensional message-type system (see also Karrer and Newman \cite{karnewcom}).

In our endeavour to understand the relationship between these two formalisms, we have shown that the pair-based moment closure formalism is applicable to arbitrary recovery processes and, for non-tree networks, provides a lower bound on $\la S_i \ra$ $\forall i$ - for tree networks the representation is exact. On the other hand, we have shown that the message passing formalism is applicable to arbitrary finite networks, where the contact and/or recovery processes are pair-specific or individual-specific, and can incorporate any pure initial system state including vaccinated individuals - or any mixed initial system state where the states of individuals are independent.

\appendix

\section{APPENDIX A}

For the case where $y_i=0$ $\forall i \in V$, the expected outcome when a single initial infected is seeded uniformly at random (in a tree network) can be computed via the following methods: 1) solve the system N times with each individual in turn as the single initial infected, and then average. This would be an exact but relatively time-consuming approach. 2) increase $z$ towards 1 such that the ratio between the probability of there being one initial infected to the probability of there being more than one becomes large. We are then left with a sum of two terms, one corresponding to zero initial infecteds (contributing nothing to the time series) and the other to a single initial infected seeded uniformly at random. Thus, dividing the resulting time series (expected number infected) by the probability of having at least one initial infected, i.e. $1-z^N$, approximates the desired result. We have achieved considerable success with this second approach in our numerical computations.

\section{APPENDIX B}

Setting $t'=t-\tau$ and applying Leibniz's rule, the time derivative of Eq. \ref{H1} can be written:
\begin{align*}  \nonumber
\dot{H^{i \leftarrow j}(t)}&= -f_{ij}(0) \left( 1 - y_j - z_j\Phi^j_i(t) \right)  \\  \nonumber
& - \int_0^t \frac{\partial f_{ij}(t-t')}{\partial t} \left(1-y_j-z_j\Phi^j_i(t')\right)  \mbox{d}t',  
\tag{B1}
\end{align*}
and by then setting $f_{ij}(\tau)= T_{ij}e^{-T_{ij}\tau } \int_\tau^\infty r_j(\tau') \mbox{d} \tau'$, for Poisson contact processes, Eq. \ref{genmess} is obtained. Now, substituting from Eqs. \ref{phi1}, \ref{singles} and \ref{pair2}, and setting $g_{ij}(\tau)=r_j(\tau)e^{-T_{ij}\tau}$, for Poisson contact processes, we get:
\begin{align*}
\dot{H^{i\leftarrow j}(t)}&= -T_{ij}\frac{ \langle S_i I_j \rangle H^{i \leftarrow j}(t) }{ \langle S_i \rangle }.
\tag{B2}
\end{align*}
It is thus straightforward to derive the pair-based system represented by Eq. \ref{genpair} by computing the time derivatives of the right-hand-sides of Eqs. \ref{singles}, \ref{pair1} and \ref{pair2}, and then using these same equations to express the derivatives in terms of individual states and pair states. In the case of exponentially distributed infectious periods the pair-based moment closure equations of Sharkey et al. \cite{sharkeyA} emerge.
\section{APPENDIX C}

 Let $F^{i\leftarrow j}_0(t)= H^{i\leftarrow j}(t) \in (0,1]$ $(\forall i,j : j \in N_i)$, and define an iterative process (as in \cite{karnew}):
\begin{equation}
F^{i\leftarrow j}_{n+1}(t) = 1- \int_0^t f_{ij}(\tau)\Big[1-y_j- z_j \prod_{\substack{k \in N_j \\  k \neq i}}F^{j \leftarrow k}_n (t-\tau)\Big] \mbox{d} \tau. 
\tag{C1} 
\end{equation} 
From Eq. \ref{ineqH} we have:
\begin{equation}
F^{i\leftarrow j}_{1}(t) \le F^{i\leftarrow j}_{0}(t),
\tag{C2}
\end{equation}
and since this is true in all cases, we have:
\begin{equation}
F^{i\leftarrow j}_1(t) \ge 1- \int_0^t f_{ij}(\tau)\Big[1-y_j- z_j \prod_{\substack{k \in N_j \\  k \neq i}}F^{j \leftarrow k}_1 (t-\tau)\Big] \mbox{d} \tau, 
\tag{C3}
\end{equation}
and so in general
\begin{equation}
F^{i\leftarrow j}_{n+1}(t) \le F^{i\leftarrow j}_{n}(t).
\tag{C4}
\end{equation}

Since $F^{i\leftarrow j}_n(t)$ is bounded below by $1-\int_0^t f_{ij}(\tau) \mbox{d} \tau$ \cite{karnew}, this iterative procedure must converge from above such that
\begin{equation}
F^{i\leftarrow j}_n(t) \to F^{i\leftarrow j}(t) \le H^{i\leftarrow j}(t)  \qquad \mbox{as } n \to \infty.
\tag{C5}
\end{equation}

\end{document}